# Pulse duration dependence of material response in ultrafast laser-induced surface-penetrating nanovoids in fused silica

Guodong Zhang[1], Na Li[1], Hao Zhang[1], Huaiyi Wang[1], Jinlong Xu[1], Jiang Wang[1], Jing Wang[2,*], Dandan Hui[3,*], Yuxi Fu[3], Guanghua Cheng[1,*]

[1] School of Artificial Intelligence, OPtics and ElectroNics (iOPEN), Northwestern Polytechnical University, Xi'an 710072, China
[2] Science and Technology on Thermostructure Compsite Materials Laboratory, Northwestern Polytechnical University, Xi'an, 710072, China
[3] State Key Laboratory of Ultrafast Optical Science and Technology, Xi'an Institute of Optics and Precision Mechanics, Chinese Academy of Sciences, Xi'an 710119, China
*Corresponding author: wangjing1@nwpu.edu.cn, huidandan@opt.ac.cn, guanghuacheng@nwpu.edu.cn

The focused ultrafast laser, with its ability to initiate nonlinear absorption in transparent materials, has emerged as one of the most effective approaches for micro-nano processing. In this study, we carried out research on the processing of high-aspect-ratio nanovoids on fused silica by using the single-pulse ultrafast Bessel beam. The thermodynamic response behaviors of the materials on surface and deep inside are found to exhibit pronounced disparities with the variation in laser pulse duration. As the pulse duration increases from 0.2 ps to 9.0 ps, the intensity of material ablation on silica surface exhibits a gradually decreasing trend, while for the void formation deep inside silica, the void diameter exhibits a trend of initial increase followed by decrease. In particular, no nanovoids are even induced deep inside when the pulse duration is 0.2 ps. The mechanism causing such differences is discussed and considered to be related to the peak intensity, group velocity dispersion, and plasma defocusing. By covering a polymer film on silica surface to influence the energy deposition, the thermomechanical response behaviors of the materials to laser pulse duration are modulated, and the material sputtering on nanovoid opening is suppressed. On this basis, surface-penetrating nanovoid arrays are fabricated on a 2-mm-thick silica sample using 2 ps Bessel beam. Given the nanovoid diameter of approximately 150 nm, the aspect ratio of the nanovoids on fused silica sample exceeds 13000:1. This outcome creates significant possibilities for the stealth dicing and processing of 3D photonic crystals, optical integrated devices, and nanofluidics.

## 1. INTRODUCTION

The ultrafast laser has emerged as a powerful tool for micro-nano processing in recent years[1-3]. Due to the short timescale for laser-mater interaction and high peak intensity, the focused ultrafast laser can trigger electron excitation of materials without inducing large-scale heat-affected zones, giving them significant advantages in improving the spatial resolution of material processing. Compared with nanosecond lasers, utilizing femtosecond lasers for processing enables the reduction of the heat-affected zone from the order of tens of micrometers to the sub-micrometer level. More importantly, the focused ultrafast laser can directly initiate the nonlinear absorption of transparent material, inducing localized electronic and thermomechanical relaxations, and therefore modifications on physical and chemical properties of material[4-6]. Surface structuring and three-dimensional selective processing of transparent samples can be realized conveniently[7], exhibiting remarkable application value in processing optical storage[8], optical microcavities[9, 10], photonic crystals[11], microfluidic chips[12], and other devices[13, 14].

However, during the ultrafast laser irradiation, there exist a complex interplay among factors such as material properties, laser parameters, and the external environment including temperature and gas atmosphere. These factors are coupled and exert unique influences on the laser-matter interaction processes, presenting both challenges and opportunities for regulating the ultrafast laser modification. Du et al. reported the influence of pulse duration on laser-induced breakdown inside fused silica in a single-shot regime[15]. When the laser pulse duration is less than ~1 ps, the breakdown threshold of fused silica was found not to meet the scaling rule of $\sqrt{\tau}$ but $\tau^{-1}$. This remarkable feature was interpreted to be related to the competition of impact ionization and nonlinear photoionization[16]. A similar phenomenon also occurs in the fabrication of nanovoids within fused silica in the single-shot regime. As reported by Razvan et al.[17], a shorter pulse duration does not necessarily promote more favorable conditions for nanovoids formation. When it comes to sample surface processing, the abrupt interface change engenders distinct scenarios in laser energy deposition and thermomechanical field relaxation[18-22] compared with internal processing. The ejection and redeposition of materials, coupled with shock/stress waves, substantially exacerbate the complexity of morphological modification at the sample interface. The outcomes of laser surface structuring exhibit a pronounced sensitivity to the location of the laser focus. Slightly altering the position of the laser focus in the vicinity of sample surface can induce diverse morphological features such as ablated craters, smooth bulges, and holes[23]. In this context, employing ultrafast Bessel beams for surface-penetration processing of transparent materials may lead to intriguing phenomena. Since the focused ultrafast Bessel beam can provide one-dimensional focal field with a wavelength-scale diameter and millimeter-scale focal depth (non-diffractive length)[24-26], it enables concurrently stimulating the materials both on the surface and deep inside sample in the single-shot regime. The discrepancies in energy absorption and thermomechanical relaxation between the sample surface and interior may exert significant influences on the modification characteristics along the longitudinal direction.

To investigate the characteristics of material responses in ultrafast Bessel beam processing and the disparities in laser acting on the surface and the interior of materials, we carry out comparative studies in this work. The non-diffractive zone of ultrafast Bessel beam is respectively placed on the front surface and inside silica samples to trigger energy deposition along a high-aspect ratio nano-channel. Through the morphological characterization and numerical simulations, the material responses to laser irradiation and the influences of pulse duration are analyzed. It is found that there exists a strong and steep enhancement in energy deposition at the silica front surface layer, which influences the morphological responses of surface material to chirped ultrafast laser pulse. Leveraging the regulation of material response behavior, we have successfully fabricated nanovoids with an ultra-high aspect ratio of 13000:1, which traverse the 2-mm-thick silica sample.

## 2. EXPERIMENT

A 1030 nm ultrafast laser (Pharos, Light Conversion) with tunable pulse duration (0.29 ps~10.0 ps) and repetition rate is used in the experiment. An axicon with a base angle of 1° is utilized to realize the conversion from Gaussian beam to non-diffractive Bessel beam. A 4f telescopic arrangement (consisting of lens1 and lens2) with magnification of about 45 is placed after the axicon, as shown in Fig. 1, which generates a demagnified Bessel beam with a half-cone angle of 19.5° and a non-diffractive length of about 300 μm in air (437 μm in glass). For special cases (processing array of ultra-high aspect ratio nanovoids), the axicon with a base angle of 5° in combination with a 4f system with magnification of 8.75 is used in the experiment, extending the non-diffracting length of bessel beam to be more than 2 mm. By moving the air-bearing stage (Aerotech ANT 130), the demagnified Bessel beam can be placed at any position of the silica sample which has a size of 2×10×20 mm$^3$ in the experiment. By controlling the laser repetition rate and stage moving speed, the ultrafast laser irradiation is enabled in a single-shot regime. The phase contrast microscopy (PCM), scanning electron microscopy (SEM), mechanical milling and ion-beam milling are employed to characterize the material responses to laser excitation on surface and in bulk.

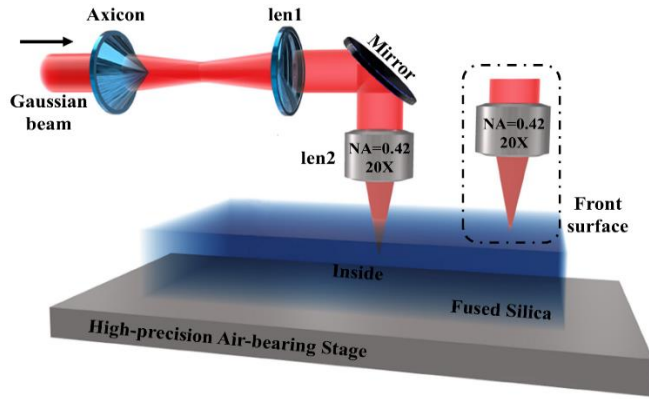

Fig. 1 Schematic diagram for the ultrafast Bessel beam inducing modification on front silica surface, rear surface, and in bulk.

## 3. RESULTS AND DISCUSSION

### a. ultrafast Bessel beam processing inside silica

In order to investigate the scenarios of ultrafast Bessel beam acting on the interior of silica, the non-diffractive zone of Bessel beam is fully placed inside the sample by moving the air-bearing stage. The laser pulse energy is fixed at 10 μJ, while the pulse duration varies from 0.2 ps to 9.0 ps. Fig. 2(a) presents the PCM images of one-dimensional laser modification after single-shot excitation. The inset curve presents the gray-level distribution along the blue dashed line. It should be noted that the increase or decrease in gray-level distribution of a PCM image corresponds to the zones where material experiences a relative decrease or increase in refractive index, respectively. Fig. 2(b) presents the SEM images of the cross-sectional view of laser modification after post- mechanical and ion-beam milling process.

It can be found that two types of modification occur inside the silica during pulse duration scanning. As for the modification induced by 0.2 ps ultrafast Bessel beam, the local material is found to experience a slight increase in refractive index, known as the Type I modification, indicating a soft structural change. It should be noted here that no nanovoid is induced under this condition, and the shallow pits in the SEM image result from the slightly higher material removal efficiency at the laser-irradiated center during ion-beam milling. When the pulse duration is increased to be 1.0 ps, the material is found to experience an apparent decrease in refractive index, known as the Type II modification, indicating an abrupt density reduction. Similar phenomenon was observed also by Bhuyan et al.[27] with using 800 nm ultrafast laser, and was attributed to the intense thermomechanical relaxations of material under strong excitation. The corresponding SEM image in Fig. 2(b) prove that nanovoid with lateral size of ~ 100 nm, and aspect-ratio larger than 3000:1, is induced under this condition. The transition from Type I to Type II is found at 0.5 ps, and the formation of longest nanovoid is found at 1.5 ps. When the pulse duration is increased to above 1.5 ps, the uniformity and length of the modified region deteriorate, and it ultimately transforms from one-dimensional nanovoids into fragmented and discrete nanodots.

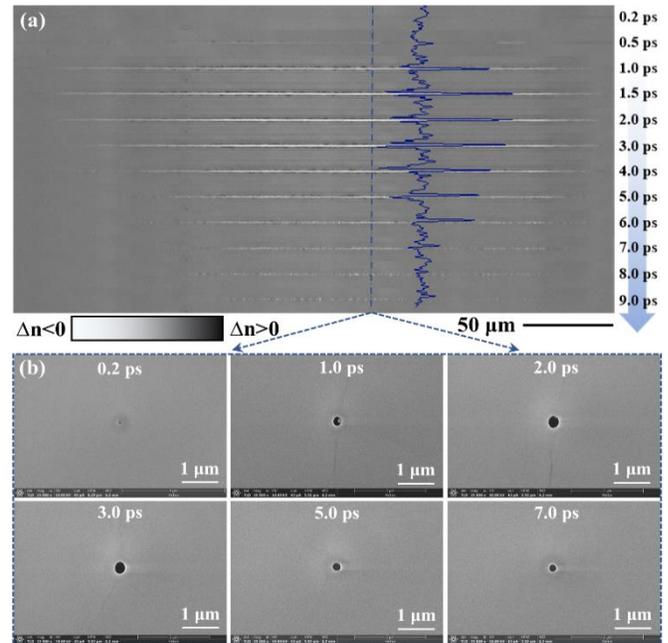

Fig. 2 PCM image of modification inside fused silica induced by single-shot ultrafast Bessel beam with fixed pulse energy of 10.0 μJ and tunable pulse duration from 0.2 ps to 9.0 ps. △n represents the refractive index change of the material after laser modification. Inset curve exhibits the relative refractive index change along the blue dashed line.

### b. ultrafast Bessel beam processing on silica surface

For comparison, we placed the non-diffracting zone of ultrafast Bessel beam on the front surface of the sample, allowing approximately one-third of the non-diffracting zone (92 μm) to be immersed below the front surface. The pulse energy used here is fixed at 10 μJ, while the pulse duration varies from 0.2 ps to 9.0 ps. By using the PCM and SEM, the morphological material responses for scenarios of ultrafast Bessel beam acting on the silica front surface are characterized. Due to the long focal depth and non-diffraction characteristics, nonlinear photoionization induced by ultrafast laser occurs not only on the surface but also deep inside silica. The PCM images in Fig. 3 (a) show that high-aspect-ratio material modification formed after single-shot Bessel beam irradiation. Regarding the nanovoid length, overall, its variation trend with pulse duration is similar to the situation when the ultrafast Bessel beam acts inside silica. However, when the pulse duration is 0.2 ps, a significant anomaly occurs. According to the scenario described in the previous section where the non-diffractive zone of the Bessel beam is fully placed inside silica, nanovoids should not form at 0.2 ps. Nevertheless, we found that while the material modification in the region deep inside silica follows Type I, near the front surface, it changes from Type I to Type II, generating a 35-μm-long nanovoid that penetrates the silica front surface.

The SEM images of silica surface morphology in Fig. 3(b) shows the surface ablation overlapping deep void initiation is triggered at the laser irradiated region, which is different from that of Gaussian beam irradiation. Intense multiple ring-shaped ejections in a fluid appearance are induced at the opening of the surface-through voids. As for the situation of 0.2 ps laser excitation, the silica nanowires with an outward radiation shape even emerge. Considering that this morphological feature is induced under the single-shot laser irradiation regime, it implies that multiple pressure waves[18] accompanying material spraying may be triggered during the thermodynamic relaxation process after laser energy deposition.

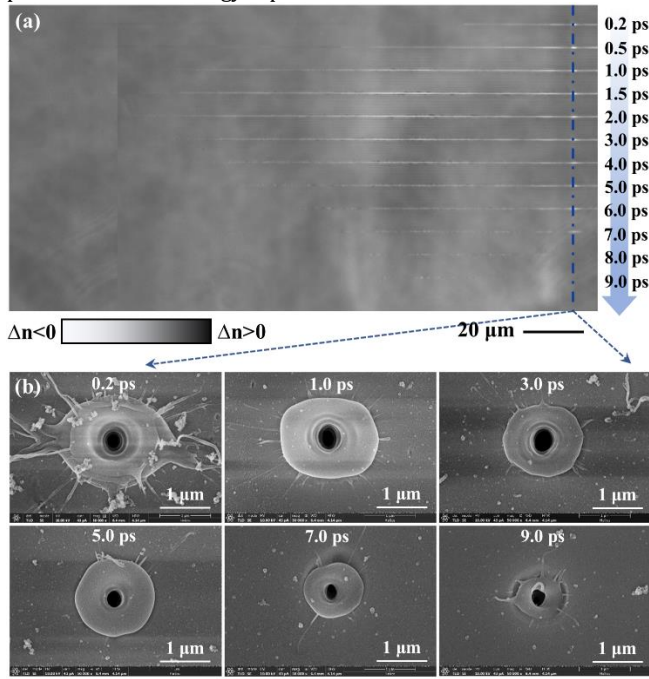

Fig. 3 (a) PCM images of silica after irradiated by single-shot ultrafast Bessel beam with fixed pulse energy of 10.0 μJ and tunable pulse duration from 0.2 ps to 9.0 ps. About one-third of the nondiffractive length is immersed below the front surface. Inset blue dashed line represents the front silica surface. (b) Corresponding SEM images of silica front surface after irradiation by single-shot bessel beam with fixed pulse energy of 10.0 μJ.

Considering the liquid ejection distance represents to some extent the material kinetic energy and therefore the maximum local pressure, one can assess the intensity of laser-induced thermomechanical relaxations and photoionization on surface from the ejection traces. In this context, the outer diameter $D_1$ of ejections elevated above silica front surface is measured and concluded as shown in Fig. 4(a). As for 0.2 ps laser excitation, the outer diameter of ejections is measured to be about 3.3 μm. When the pulse duration further increases to 9.0 ps, the outer diameter of ejections decreases monotonically to ~0.9 μm, indicating the intensity of the laser-induced thermomechanical relaxation keeps negative correlation with pulse duration. This pulse duration dependence is quite different from that obtained in ultrafast Bessel beam acting on the interior of silica, reflecting a probable difference in the photoionization or material relaxations regime.

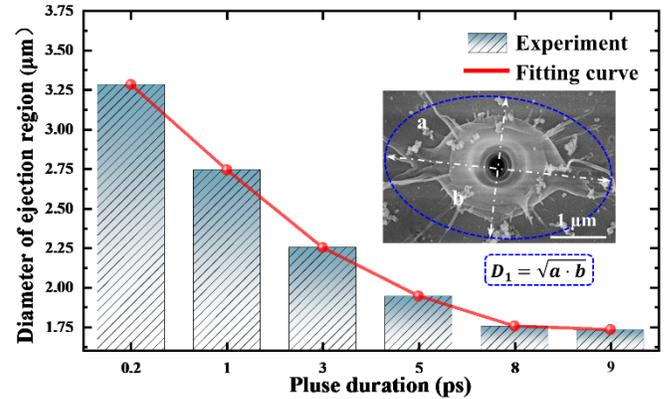

Fig. 4 SEM images of silica front surface after modified by single-shot ultrafast Bessel beam with fixed pulse energy of 10.0 μJ and tunable pulse duration from 0.2 ps to 9.0 ps. About one-third of the nondiffractive length is immersed below the front surface.

### c. simulation on the ultrafast laser-matter interaction

In order to get a deep understanding of the pulse duration dependences of laser-induced modifications on silica surface and deep inside, we developed a nonlinear propagation model analogous to the ones presented in Refs. 24 and 25 with material parameters similar to Ref. 26[28-30]. The basic processes including multiphoton ionization, avalanche ionization, free-carrier absorption, group velocity dispersion, Kerr effect and plasma defocusing were taken into account for a Gaussian pulse $\epsilon(r,z,t) = \epsilon_0 exp\left(-\frac{r^2}{\omega^2} - \frac{t^2}{\tau_p^2}\right)$ which was further modulated by the axicon and lens in phase as numerically defined by their transmittance functions. The $r$, $x$, and $t$ are the radial, axial, and time coordinate of the complex amplitude $\epsilon(r,x,t)$ of the laser field, respectively. The beam waist $\omega$ is set to be 6.8 mm, while the pulse duration $\tau_p$ is defined by the FWHM pulse duration $\tau$ as $\sqrt{\tau^2/(-2\log(1/2))}$. The Bessel beam propagation in air is assumed to be free of dispersion and nonlinear effects, and the FWHM value of central lobe is calculated to be about 1.1 μm. Consistent with the experiments, about one-third of the non-diffracting zone is placed below the silica front surface in the simulation.

Fig. 5(a-c) present the calculated distributions of laser fluence, peak excited electron density and deposited energy density for 0.2 ps ultrafast Bessel beam irradiating silica front surface. It is apparent that intense photoionization is triggered when ultrafast Bessel beam propagates into the fused silica. Moreover, the photoionization occurrs not only at the central lobe but also at the side lobes, resulting a larger excited region in longitudinal section, i.e. a larger region for energy deposition. The calculation results for the scenario of 2.0 ps ultrafast Bessel beam irradiation are shown in Fig. 5(d-f). Due to the relatively low peak intensity and weak plasma defocusing, the laser energy deposition has a more concentrated region in the longitudinal section. As a result, the deposited energy density inside silica for 2.0 ps laser irradiation is even higher than that for 0.2 ps laser irradiation. The curves of laser fluence, peak excited electron density and deposited energy density along the axis of Bessel beam for 0.2 ps and 2.0 ps in Fig. 5(g-i) demonstrate well the above abnormal results. Obviously, deep inside silica, the laser fluence, peak excited electron density, and deposited energy density for the 2.0 ps irradiation laser are significantly higher than those for the 0.2 ps laser irradiation. The maximum value of the deposited energy density inside silica for the 2.0 ps is ~9×10⁹ J/m³. With using the heat capacity formula $\Delta T = U/(c \cdot \rho)$, the temperature

at the central excited region is roughly estimated to be about 4500 K. This indicates to some extent that strong thermomechanical relaxations including expansion, phase transition, and cavitation can be triggered within the localized region. In contrast, the maximum value of the deposited energy density inside silica for 0.2 ps is ~$5\times10^9$ J/m$^3$, the estimated temperature is only about 2500 K. This provides a compelling explanation for the absence of nanovoids when 0.2 ps ultrafast Bessel beam irradiates deep inside silica.

In addition, we find that significant and sharp enhancements exist in the laser fluence, peak excited electron density, and deposited energy density at the silica front surface layer (with a thickness of ~100 nm). Concurrently, strong fluctuations occur in the underlying layer with a thickness on the order of ~1 μm. Notably, whether in the front surface layer or the region where fluctuations occur, the deposited energy density for 0.2 ps laser irradiation is significantly higher than that for 2.0 ps laser irradiation. By analyzing the characteristics of the nonlinear photoionization processes, the phenomena are inferred to be related with the group velocity dispersion and plasma defocusing which possess spatial hysteresis in suppressing the photoionization, i.e. the energy deposition. Given the energy deposition determines the consequent thermomechanical relaxations and morphological responses of material, these sharp enhancements and fluctuations near silica surface are proposed to be the key reason for the variation of morphological responses of material with depth, nevertheless, the possible change of physical strength of material near surface may also perform a role.

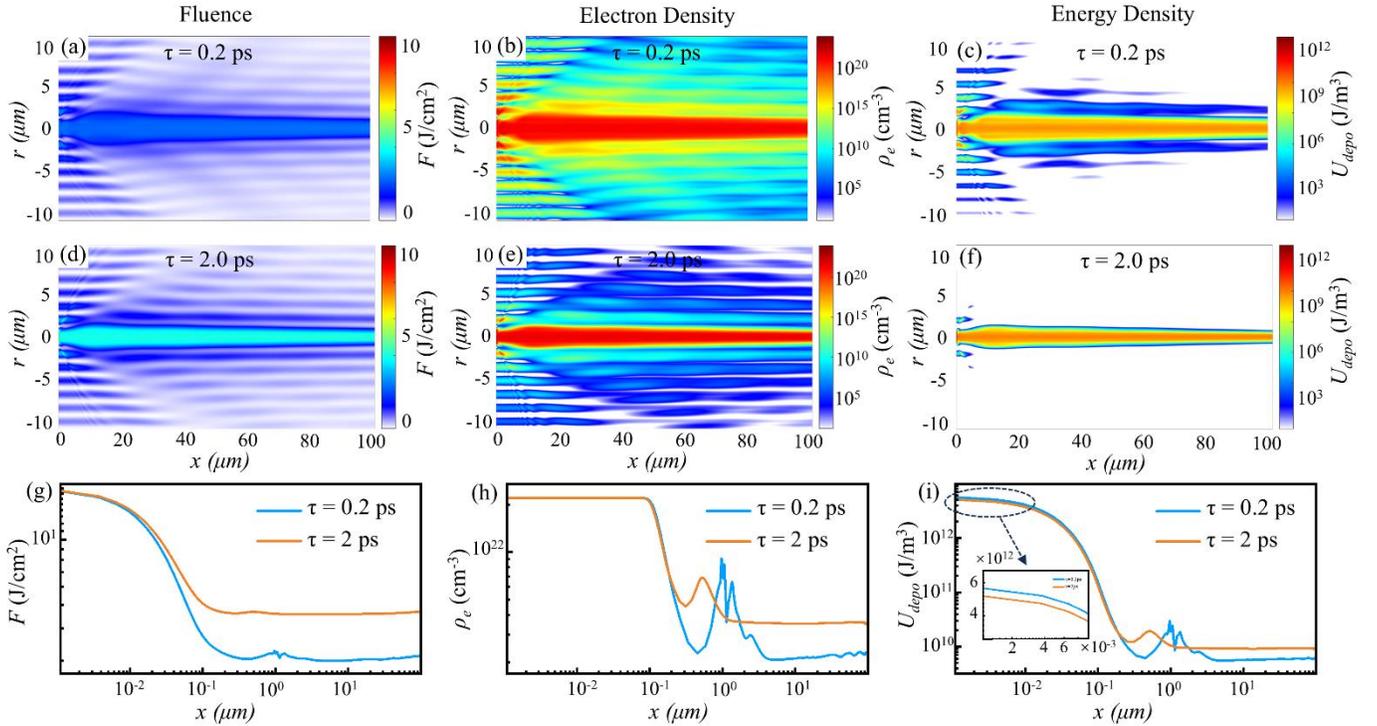

Fig. 5 Calculated distributions of laser fluence(a, d), peak excited electron density(b, e) and deposited energy density(c, f) for 0.2 ps and 2.0 ps ultrafast Bessel beam irradiating silica front surface. About one-third of the nondiffractive zone is immersed below the front surface. The laser fluence, peak electron density and deposited energy density along the axis of Bessel beam ($r = 0$) are respectively plotted in (g-i). The silica upper surface locates at z = 0 μm.

## d. regulation of material responses to laser irradiation on the surface

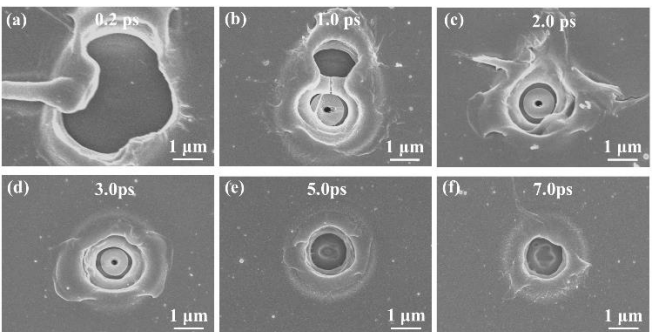

Fig. 6 Surface morphology of the silica covered with PMMA after irradiation by single-shot Bessel beam with a fixed pulse energy of 10.0 μJ. About one-third of the nondiffractive zone is immersed below the front surface.

To explore the controllability of the response behavior of surface materials to lasers, we deposited a film of polymethyl methacrylate (PMMA) with a thickness of ~ 400 nm on the surface of silica sample, aiming to influence the laser energy deposition process as well as the thermomechanical relaxation of the material. Compared with fused silica, PMMA has a smaller energy band gap (~4.4 eV) and a lower melting point (~180 K)[31]. The non-diffracting zone of ultrafast Bessel beam was placed on the sample front surface, allowing approximately one-third of the non-diffracting zone (92 μm) to be immersed below the front surface. The pulse energy used here is fixed at 10 μJ, while the pulse duration varies from 0.2 ps to 7.0 ps. The SEM images in Fig. 6 present the surface morphologies after single-shot laser irradiation.

It is evident that the PMMA film within the laser-irradiated region underwent intense ablation and spraying, exhibiting distinct morphological characteristics from the underlying silica surface. For pulse duration of 0.2 ps, the ablated area of the PMMA film is the largest, while there are no apparent modification signs or nanovoids observed on the underlying silica surface. As the pulse duration further increases, the ablated area of the PMMA film displays a tendency of gradual decrease. However, distinctively, when the pulse duration reaches above 1.0 ps, nanovoids and highly regular donut-shaped ejecta begin to appear on the silica surface. Therefore, it can be concluded that the variation trend of the response behavior of the silica surface material with respect to the pulse duration becomes similar to the scenario when the ultrafast Bessel beam acts on the deep interior of silica. The modulation mechanism of the deposited PMMA film is considered as follows: it enables the surface enhancements above shift from the silica surface layer to the PMMA film, thereby narrowing the gap in the deposited energy density between the silica surface and the deep interior.

### e. Fabrication of ultrahigh-aspect-ratio nanovoids

Based on the above work, we carried out the fabrication of surface-penetrating nanovoid arrays on a 2-mm-thick silica sample using 2ps ultrafast Bessel beam. The period of the nanovoids is 2.5 μm. Fig. 7 (a) present the fabricated silica sample illuminated by white light, which is essentially a two-dimensional volume grating and exhibits obvious optical diffraction characteristics. Enabled by the surface modulation, the openings of the nanovoids do not show random material sputtering characteristics, but present a regular donut-like morphology, as shown in Fig. 7 (b). By cleaving the sample in different directions, we obtained the lateral and longitudinal information of the nanovoids with the help of the SEM. Images at the right side of Fig. 7 (b) show the longitudinal cross-section view of the nanovoids. It can be measured that the diameter of the nanovoids is approximately 150 nm. Considering that the depth of the nanovoids reaches 2 mm, the aspect ratio of the nanovoids exceeds 13000:1. It is noteworthy that there is an irregular serrated boundary outside the walls of the nanovoids, with a left-right spacing of 821 nm. We attribute this to the nanocracks formed outside the nanovoids during cavitation.

The lateral cross-section view of the nanovoid array is presented in Fig. 7 (c), while its local enlarged views are shown in Fig. 7 (d) and 7 (e). The cross-section of the nanovoid exhibits a high degree of circular symmetry, and the measured diameter is consistent with that measured in the longitudinal section. It can be observed that there are radially distributed nanocracks on the outer side of each nanovoid. These nanocracks do not extend to the nanovoid walls but are distributed in the annular region outside the nanovoids, which we refer to as Region 2. The outer diameter of Region 2 can reach approximately 2.5 μm. Since Region 2 is predominantly located outside the central core of the Bessel beam, we attribute nanocrack formation to the local high pressure from the thermal-expansion-induced outward release of material within the beam's main lobe. There are no nanocracks in the region between Region 2 and the nanovoids, which we call Region 1. The outer diameter of Region 1 is approximately 800 nm. The value is consistent with the left-right spacing of the serrated boundary in the longitudinal section. Since Region 1 lies within the central core of the Bessel beam, the material in this region undergoes a melting and subsequent re-solidification process after energy deposition. We believe that this process is the primary reason for the absence of nanocracks in Region 1. A detailed discussion on this point was carried out in our previous work[32]. The direct writing of ultra-high aspect-ratio nanovoids without intense surface damage offers the potential for the fabrication of functional devices (e.g., nanosieves, photonic crystals, volume grating) , as well as for the ultra-precision cutting of materials.

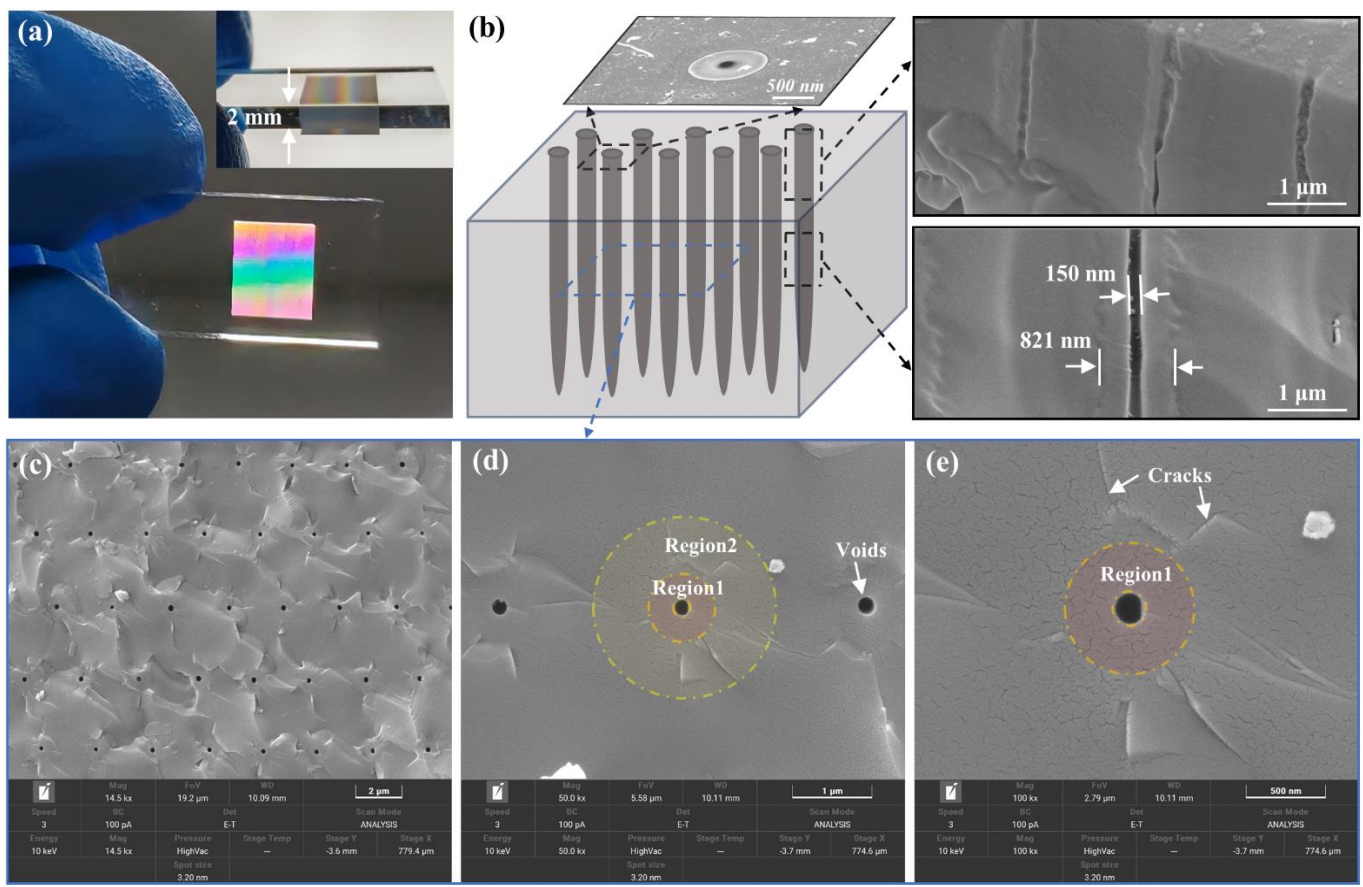

Fig. 7. (a) The 2 mm-thick sample with an embedded array of ultra-high aspect ratio nanovoids fabricated using 2ps ultrafast Bessel beam. (b) SEM images of the longitudinal section of nanovoid and the nanovoid opening on sample front surface, (c-e) Cross section of the nanovoid array within the sample and its magnified view.

## 4. CONCLUSION

In summary, we investigated the characteristics of ultrafast Bessel beams in the fabrication of nanovoids in fused silica. Experimental findings revealed that when inducing high-aspect-ratio nanovoids, materials on the sample surface and deep inside exhibited distinct thermodynamic responses to laser irradiations of different pulse durations. Through simulations of the nonlinear propagation of ultrafast Bessel beams in fused silica, we discovered that significant and sharp enhancements existed in the deposited energy density at the silica front surface layer. Overall, the deposited energy density at the silica surface is considerably higher than that inside the sample. The phenomena are inferred to be related to the group velocity dispersion and plasma defocusing which possess spatial hysteresis in suppressing photoionization, i.e. the energy deposition. This outcome reasonably explains the stronger thermomechanical response at the nanovoid opening. As the pulse duration varied, the deposited energy density on the silica surface and deep inside exhibited different response behaviors. In the front surface layer, the deposited energy density for 0.2 ps laser irradiation was significantly higher than that for 2.0 ps laser irradiation, whereas inside the sample, the deposited energy density for 0.2 ps was significantly lower. This provides a compelling explanation for the absence of nanovoids when 0.2 ps ultrafast Bessel beam irradiates deep inside silica. By covering a polymer film on the silica surface to influence the energy deposition, the thermomechanical response behaviors of the materials to laser pulse duration were modulated. Based on this, surface-penetrating nanovoids with a record aspect ratio of 13,000:1 were fabricated on a 2-mm-thick silica sample using 2 ps Bessel beam, with suppressed material sputtering on the nanovoid opening. These results hold great promise for the fabrication of functional devices (e.g., nanosieves, photonic crystals, volume grating) , as well as for the ultra-precision cutting of materials.

**Acknowledgments**

The research work was supported by the National Key R&D Program of China (2024YFB4609200), National Natural Science Foundation of China (U24A20125), Natural Science Basic Research Program of Shaanxi Province (2022JQ-648, 2025JC-YBMS-477), Creative Research Foundation of the Science and Technology on Thermostructural Composite Materials Laboratory (JCKYS 2024 607004).

**Author contribution**

Guodong Zhang: Experimental Investigation, Conceptualization, Data curation, Writing – original draft, Supervision. Na Li: Experimental Investigation, Data curation. Hao Zhang: Simulation, Formal analysis, Writing – review & editing. Huaiyi Wang: Experimental Investigation, Data curation. Jing Wang: Writing – review & editing, Formal analysis, Funding acquisition. Dandan Hui: Writing – review & editing, Formal analysis, Funding acquisition. Yuxi Fu: Writing – review & editing, Formal analysis. Guanghua Cheng: Resources, Funding acquisition, Writing – review & editing, Supervision.

**Conflict of interest**

The authors declare no conflict of interest.

**Data availability.**

Data underlying the results presented in this paper are available in.